\documentstyle[preprint,aps,psfig]{revtex}

\input{psfig}
\def\be{\begin{equation}}
\def\ee{\end{equation}}
\def\bea{\begin{eqnarray}}
\def\eea{\end{eqnarray}}
\begin{document}

\title{Global observables and secondary interactions 
in central Au+Au reactions at $\sqrt{s}=200$~AGeV}

\author{M.J.~Bleicher${}^{a \xi}$\thanks{E-mail: 
bleicher@nta2.lbl.gov}, 
S.A.~Bass${}^{b \xi}$, L.V. Bravina${}^{cd}$, W. Greiner${}^{e}$, 
S. Soff${}^{f}$, \\
H.~St\"ocker${}^{e}$, N. Xu${}^{a}$, E.E. Zabrodin${}^{cd}$}

\address{${}^a$ Nuclear Science Division,
Lawrence Berkeley Laboratory, Berkeley, CA 94720, U.S.A.\\}

\address{${}^b$Department of Physics, Duke University,
        Durham, N.C. 27708-0305, USA}

\address{${}^c$Institut f\"ur Theoretische Physik, Universit\"at T\"ubingen,
72076 T\"ubingen, Germany}

\address{${}^d$Institut for Nuclear Physics, Moscow State University, 119899
Moscow, Russia}  

\address{${}^e$Institut f\"ur
Theoretische Physik,  Goethe-Universit\"at,
60054 Frankfurt am Main, Germany}

\address{${}^f$ Gesellschaft f\"ur Schwerionenforschung, Darmstadt, Germany} 

\footnotetext{${}^\xi$ Feodor Lynen Fellow of the Alexander v. 
Humboldt Foundation}

%%%%%%%%%%%%%%%%%%%%%%%%%%%%%%%%%%%%%%%%%%%%%%%%%%%%%%%%%%%%%%
% You may repeat \author \address as often as necessary      %
%%%%%%%%%%%%%%%%%%%%%%%%%%%%%%%%%%%%%%%%%%%%%%%%%%%%%%%%%%%%%%

\maketitle
\begin{abstract}
The Ultra-relativistic Quantum Molecular Dynamics model (UrQMD) is used
to study global observables in central reactions of Au+Au at
$\sqrt{s}=200$~AGeV (RHIC). Strong stopping governed by massive particle
production is predicted if secondary interactions are taken into account. 
The underlying string dynamics and the early hadronic decoupling implies only
small transverse expansion rates. However, rescattering with mesons is found to
act as a source of pressure leading to additional flow of baryons and kaons, 
while cooling down pions. 
\vfill 
\noindent
LBNL-preprint: LBNL-44599
\end{abstract}

\newpage
%\vspace*{3cm}
One of the major goals of the relativistic heavy ion collider (RHIC) at
Brookhaven National Laboratory is to explore the phase diagram of hot 
and dense matter near the quark gluon
plasma (QGP) phase transition. 
The QGP is a  state in which the individual hadrons dissolve into a
gas of free (or almost free) quarks and gluons in strongly compressed
and hot matter (for recent reviews on the topic, we refer to
\cite{qgprev,mu1}). 
The achievable energy- and baryon densities
sensitively depend on the extend to which the nuclei are stopped during
penetration; they also depend on mass number and bombarding energy.

Earlier RHIC estimates have been performed assuming boost-invariant
hydrodynamics \cite{Bj,clare86a,rischke96a,dumitru,hung98a} 
and pQCD (Regge theory) motivated model \cite{geiger,hijing}: baryons are 
concentrated at projectile and target rapidity
separated by a large region which is baryon free (in position and
momentum space), i.e. the nuclei are transparent. The region between
them is filled by the color fields which materialize, developing a
plateau in the mesons' rapidity distribution. This scenario is supported
experimentally for pp- and p$\bar{\rm p}$-collisions at collider
energies. It is the aim of the present work to examine whether this
remains true also for the collision of large nuclei. From lower energy
nucleus-nucleus collisions we know that only a small fraction of the
total number of collisions takes place at the full incident energy while
most of them take place at much lower energies. In fact, transport 
model studies show a fair amount of stopping at the RHIC energy with strong
transverse expansion \cite{schoenfeld,monreal} indicating that the
collision of two nuclei is more than just the superposition of 
''A$\times$A'' nucleon collisions at
the same energy (i.e. that secondary interactions are very important at all
investigated energies).

As a tool for our investigation of heavy ion reactions at RHIC the 
Ultra-relativistic Quantum Molecular Dynamics model (UrQMD 1.2) is 
applied \cite{urqmd}. 

Similar to the RQMD model \cite{schoenfeld,sorge}, UrQMD is a microscopic transport
 approach based on the covariant propagation of
constituent quarks and diquarks accompanied by mesonic and baryonic 
degrees of freedom.
It simulates multiple interactions of ingoing and newly produced 
particles, the excitation
and fragmentation of color strings and the formation and decay of
hadronic resonances. 
At RHIC energies, the treatment of subhadronic degrees of freedom is
of major importance.
In the UrQMD model, these degrees of freedom enter via
the introduction of a formation time for hadrons produced in the 
fragmentation of strings \cite{andersson87a,andersson87b,sjoestrand94a}.
The leading hadrons of the fragmenting strings contain the valence-quarks 
of the original excited hadron. In UrQMD they are allowed to
interact even during their formation time, with a reduced cross section
defined by the additive quark model, 
thus accounting for the original valence quarks contained in that
hadron \cite{urqmd}. Those leading hadrons therefore represent a simplified
picture of the leading (di)quarks of the fragmenting string. 
Newly produced (di)quarks do, in the present model, not 
interact until they have coalesced into hadrons -- however,
they contribute to the energy density of the system.
A more advanced treatment of the partonic degrees of freedom during the
formation time ought to include soft and hard parton scattering
\cite{geiger} and the explicit time-dependence of the color interaction
between the expanding quantum wave-packets \cite{gerland98a}.
However, such an improved treatment of the internal hadron dynamics has
not been implemented for light quarks into the present model.
For further details about the UrQMD model, 
the reader is referred to Ref. \cite{urqmd}.

The UrQMD model has been applied successfully to
explore heavy ion reactions from AGS energies (E$_{\rm lab}=1-10$~AGeV) 
up to  the full CERN-SPS energy (E$_{\rm lab}=160$~AGeV). 
This includes detailed 
studies of thermalization \cite{bravina}, particle 
abundancies and spectra \cite{bl1}, strangeness production \cite{ssoff},
photonic and leptonic probes \cite{dumi} , $J/\Psi$'s \cite{jpsi} and 
event-by-event fluctuations \cite{bl2}.

Let us tackle directly the relevant questions prompted by the
start-up of RHIC:
\begin{itemize}
\item
Can string models like UrQMD be applied to AA reactions at RHIC energies?
\item
Is baryonic stopping achieved at RHIC?
\item
How many particles will be produced?
\item
Will secondary interactions modify observables?
\end{itemize}

The increasing importance of perturbative QCD effects
(hard scattering)\cite{geiger,hijing,kwerner} and coherent parton 
dynamics \cite{mclerran94} has lead to the speculations that transport models
with string dynamics will fail to describe heavy ion collisions above a certain
center of mass energy.
Indeed, todays transport models are based on a probabilistic phase 
space approach, even in the earliest stage of the reaction. 
In this stage at RHIC energies, the protons and neutrons of the colliding 
nuclei should be described by coherent parton wave 
functions and should be modelled 
as such \cite{mclerran94}. However, after
initial parton or string production has taken place in the 
first 0.5 fm/c \cite{eskola94a}, this coherence is lost and the UrQMD
ansatz may be applicable.

To study the pQCD-induced effects 
it has been suggested to use the Parton Cascade Model (PCM/VNI) \cite{geiger} 
to simulate the dynamics of the hot and dense region of heavy ion reactions. 
However, the interplay of hard vs. soft physics (early stage vs. late
stage of the collision) allows use of these models only 
in the very early stage of the reaction. Recently 
it was shown that the large amount of non-perturbative parton interactions 
at SPS and RHIC energies imposes severe limitations to the applicability of
such an approach \cite{bass9908014,mu1}. 

In fact, higher twist phenomena seem to play an important role at 
RHIC energies, making leading order (and next-to leading order) 
perturbative QCD (pQCD) calculations questionable for
the study of dense matter \cite{bass9908014,stein}. 
It has been argued \cite{M1,S1}, that for subsequent ($t> 1$~fm/c) 
collision stages, the
use of phenomenological approaches to investigate the collision
dynamics is inevitable, especially when the system
becomes relatively dilute and secondary collisions occur at moderate energies.

It is not known a priori at RHIC energies,
whether pQCD effects (presumably taking place at the early stage of the 
collision, $t \sim 1$~fm/c) or the hadronic
rescatterings dominate the evolution of the system and 
the hadronic spectra measured by the experiments after freeze-out.
Models like UrQMD \cite{urqmd} or RQMD \cite{sorge} 
can help to identify in the observables signals from 
different (early/late) stages of the collision dynamics.

Let us investigate  UrQMD predictions at increasing 
center of mass energies for light-ion and proton-proton reactions.
UrQMD calculations to rapidity distributions 
for He+He at $\sqrt{s}=31$~GeV (ISR)
yield good agreement between model and data \cite{urqmd}. 
If the energy is increased  further, $pp$ interactions from UrQMD start 
to deviate  from pQCD motivated extrapolations 
by 35\% at $\sqrt{s}=200$~GeV. This deviation is consistent with early 
attempts made in the RQMD approach as discussed in ref. \cite{schoenfeld}.
It can be pinned down to multi-jet events: here the incoming hadrons do 
fragment not only into two jets (lead by the incoming quarks/diquarks) 
but also into additional
jets stemming from momentum transfer to the sea-partons of the incoming
hadrons. These additional jets result in an overall increase of particle
production from center of mass energies of $\sqrt{s}=100$~GeV upwards. 
Consequently, 
this model cannot be applied to pQCD dominated observables,
e.g. the high-momentum ($p_t > 2 $~GeV/c) part of hadronic spectra or
multi-jet related quantities.
On the other hand, as will be discussed below, only
a minor part of all elementary interactions takes place at such high energies,
thus final results in terms of particle multiplicity and
spectral shape are only moderately affected, on the order of 10\%
\cite{schoenfeld}.

\begin{figure}[t]
\vskip 0mm
\vspace{1.0cm}
\centerline{\psfig{figure=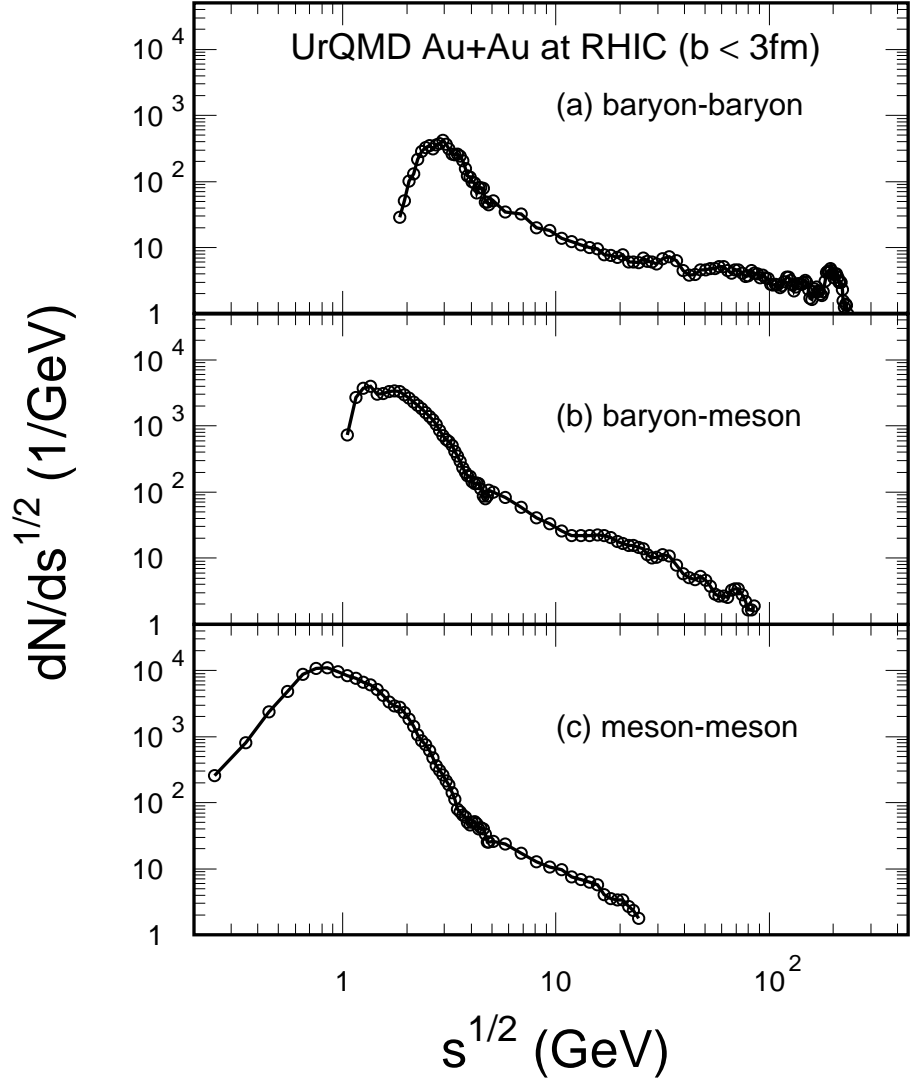,width=15cm,clip=}}
\vskip 2mm
%\vspace{2.0cm}
\caption{Collision energy spectra of baryon-baryon (a), meson-baryon
(b) and meson-meson (c) reactions in Au+Au, 
$\sqrt{s}=200$~AGeV, b$<$3~fm.\label{colldyn}}
\end{figure}

Figure \ref{colldyn} shows the $\sqrt s$-collision spectra of 
individual hadron
(quark) collisions  in Au+Au reactions at
$\sqrt{s}=200$~GeV. Fig. \ref{colldyn}a  
indicates all baryon-baryon (BB) and diquark-diquark collisions,
Fig.\ref{colldyn}b shows meson-baryon (MB) and quark-diquark reactions and 
Fig.\ref{colldyn}c describes meson-meson (MM) and quark-quark 
collisions. All
spectra are strongly decreasing towards high collision energies. However,
the initial baryon-baryon (diquark-diquark) interactions are visible as
a bump around the beam energy of $\sqrt{s}=200$~GeV (the width of this
bump is given by the Fermi-momentum multiplied by the Lorentzfactor).

One observes that the total number of collisions is 
dominated by secondary interactions.
The initial high energy collisions ($\sqrt{s}>100$~GeV) constitute 
less than 20\% of all reactions. The remaining 80\% of the reactions are 
well treatable by string physics and effective constituent quark
dynamics. The average collision energies are given by:
\be
\langle \sqrt s \rangle 
= \frac{\int {\rm d}\sqrt s \, \sqrt s \,
\frac{{\rm d}N}{{\rm d}\sqrt s}}
{\int {\rm d}\sqrt s \, \frac{{\rm d}N}{{\rm d}\sqrt s}}
\ee
resulting in $\langle \sqrt s \rangle^{\rm MM}=1.2$~GeV, $\langle \sqrt s
\rangle^{\rm MB}=2.3$~GeV and $\langle \sqrt s \rangle^{\rm BB}=8.2$~GeV.
It is interesting that the BB value is mostly driven by the initial
collisions. If only reactions below $\sqrt s = 100$~GeV are counted,
$\langle \sqrt s \rangle^{\rm BB}_{\sqrt s <100{\rm ~GeV}}=4.6$~GeV. Note 
that these moderate
collision energies are also encountered in 
'pQCD' based approaches, e.g. VNI \cite{bass9908014}. Thus  pointing 
to a strong non-perturbative component in the parton-hadron 
dynamics at RHIC energies. 

In the following two different scenarios
will be explored in order to study the influence of 
secondary interactions: UrQMD calculations with the 
full collision term included
will be contrasted by UrQMD simulations with deactivated meson-meson and
meson-baryon interactions. The following interactions
have been deactivated: Meson-meson, meson-baryon, valence quark-meson,
diquark-meson, valence quark-valence quark, valence quark-diquark (incl. 
anti-quarks and baryons).
Note that baryon-baryon, diquark-baryon and diquark-diquark collisions are
still possible. This is in contrast to first collision models: In the UrQMD
model 'without rescattering' not  only multiple
baryon-baryon interactions are allowed, but also baryon--anti-baryon 
annihilations are still possible (cf. Tab. \ref{tb1}). 

\begin{figure}[t]
\vskip 0mm
%\vspace{-1.8cm}
\centerline{\psfig{figure=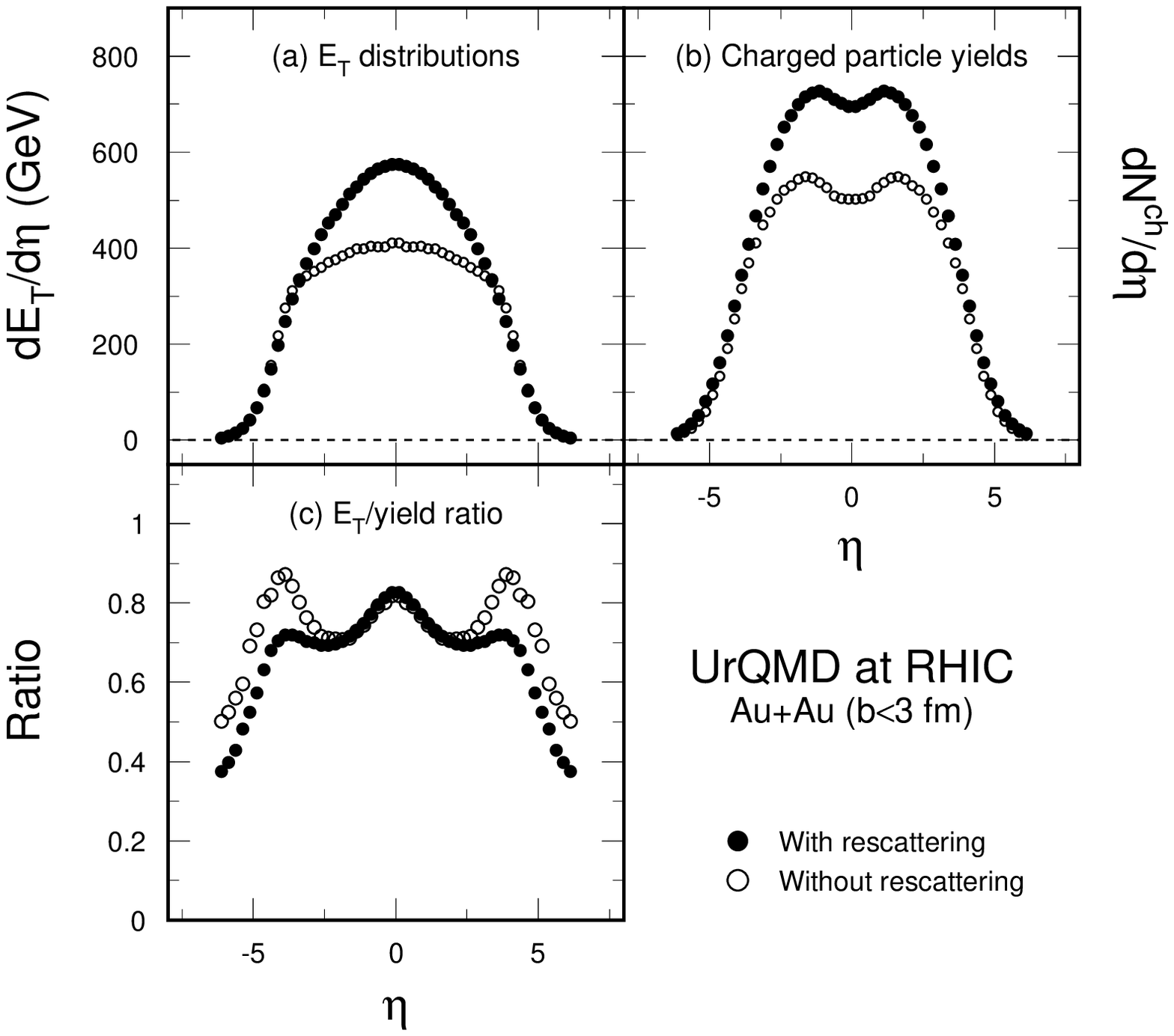,width=17cm}}
\vskip 0mm
\vspace{-.0cm}
\caption{Au+Au, $\sqrt{s}=200$~AGeV, b$<$3~fm. Full
symbols denote calculations with full rescattering. Open symbols denote 
calculations without meson-meson and meson-baryon interactions.\\ 
(a) Transverse energy distribution as a function of pseudo rapidity.\\
(b) Pseudo rapidity density of charged particles ($\pi^++\pi^-+K^++K^-$).\\
(c) Transverse energy per charged particle as a function of pseudo rapidity.
\label{et}}
\end{figure}

Let us investigate the total energy deposition in calorimeters 
in terms of the transverse energy $E_T$:
\be
E_T =  \sum \left(E_i\, {\rm sin} \, \theta_i + m_i \right) \quad,\quad 
\theta_i={\rm arctan}\frac{p_{i\perp}}{p_{i||}} \quad.
\ee
$E_i$ is the energy of particle $i$, $m_i$ is the restmass of particle $i$ - if
it is an anti-baryon, otherwise it is zero.
The $E_T$ distribution is depicted in Fig. \ref{et}a as a function 
of pseudo-rapidity $\eta$. UrQMD predicts a maximum $E_T$ of 600 GeV (with
rescattering, full symbols) and a gaussian shape of the $E_T$ distribution. 
Deactivating the secondary interactions (open symbols) results in a 
decreased  
energy deposition by 30\% and in a plateau in the transverse energy
distribution, as expected from string dynamics.  

The charged particle ($\pi^++\pi^-+K^++K^-$) yields with (full symbols) and
without (open symbols) rescattering are depicted in Fig. \ref{et}b.

Fig. \ref{et}c shows the $E_T$ per charged particle as a function of 
pseudo-rapidity. At the central region the calculation with rescattering (full
symbols) and without rescattering (open symbols) coincide. The transverse
energy per particle is 600-800 MeV. However, at larger rapidities we observe
secondary maxima in the calculation without rescattering - as shown below, they
are due to concave momentum distributions of hadrons over rapidity.

Diquark dynamics becomes the major mechanism for
the initial built-up of energy-density  \cite{weber} and particle 
production. It is therefore interesting to study the 
stopping behaviour of the present 
model. It has been claimed recently, that exotic mechanisms (e.g. baryon
junctions \cite{vance})
need to be invoked to understand the baryon number
transport at SPS and RHIC. In contrast to these approaches, the UrQMD model
mainly applies quark model cross sections to the subsequent scattering
of constituent (di-)quarks in combination with a small diquark 
breaking \cite{acap} component ($\sim$ 10\%).

\begin{figure}[t]
\vskip 0mm
\vspace{-1.0cm}
\centerline{\psfig{figure=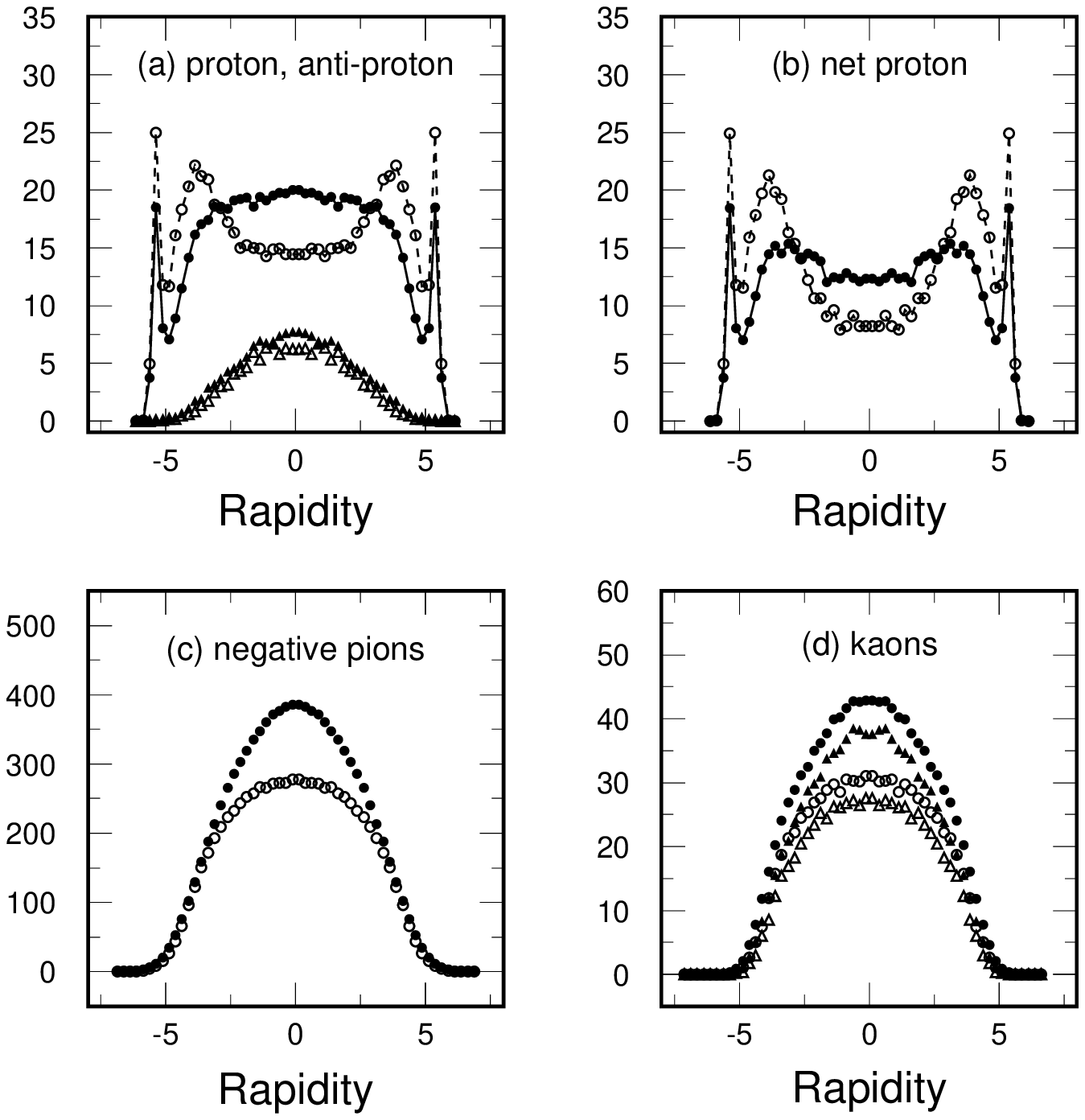,width=17cm}}
\vskip 2mm
%\vspace{-1.0cm}
\caption{Au+Au, $\sqrt{s}=200$~AGeV, b$<$3~fm. Full
symbols denote calculations with full rescattering. Open symbols denote 
calculations without meson-meson and meson-baryon interactions.\\
(a) Rapidity density of protons (circles) and anti-protons (triangles).\\
(b) Rapidity density of net-protons.\\
(c) Rapidity density of negatively charged pions.\\
(d) Rapidity density of  K$^+$ (circles) and K$^-$ (triangles).
\label{rap}}
\end{figure}

Figure \ref{rap}a shows the rapidity spectra of protons (circles)
and anti-protons (triangles) in central (b$<$3~fm) 
Au+Au reactions at $\sqrt{s}=200$~AGeV. Full symbols denote
calculations with full rescattering, whereas open symbols denote calculations 
without meson-meson and meson-baryon interactions.
The proton distribution (in the calculation with rescattering) shows 
a plateau over rapidity with 20 protons at central rapidities. 
Without rescattering the proton distribution exhibits a 
dip at central rapidity values.
The anti-proton distribution is of Gaussian shape with a peak 
value of 8 at midrapidity. It is interesting to note that the shape of the
anti-proton distribution and their absolute yield 
stays apparently unaffected by secondary interactions. Since the 
overall particle production is strongly enhanced by
rescattering effects (as shown below), this points to a counter balance
of production and annihilation of anti-baryons.

The stopping power obtained in the full UrQMD approach is rather strong
(cf. Fig. \ref{rap}b). 
We observe  a flat net-baryon rapidity distribution, while whithout
rescattering two maxima develop
near target/projectile rapidities and a strong dip at central rapidities. The
net-proton distribution (full symbols) is shifted by approximately two units in
rapidity, resulting in 12 net-protons at midrapidity. 
Secondary scatterings are important for transporting 
baryon number from projectile and target rapidity closer to midrapidity.

Figure \ref{rap}c depicts the yields of negatively 
charged pions (neutral and positively charged pions are - on 
a 5\% level - identical in shape and number) with full 
rescattering (full symbols) and 
without meson-meson and meson-baryon interactions (open symbols) 
in Au+Au, $\sqrt{s}=200$~AGeV, b$<$3~fm.
Comparing the simulations with and 
without rescattering, a strong increase of particle production in the central
rapidity region is observed if rescattering is included. 
As a result, a Gaussian shape of the pions rapidity distribution emerges.

The kaon distribution is affected by secondary 
interactions as well, as is shown in Fig.
\ref{rap}d. The rapidity distributions of K$^+$ (circles) 
and K$^-$ (triangles) with
full rescattering (full symbols) and without secondary interactions 
(open symbols) are shown for Au+Au, $\sqrt{s}=200$~AGeV, b$<$3~fm
reactions. The overall amount of charged kaons increases by nearly 30\% 
due to rescattering
effects. However, the splitting between positively and negatively charged kaons
seems to be unaffected by meson-meson and meson-baryon interactions. 
If this is the case in collisions, a possible equilibration of 
strangeness due to a QGP (as proposed by \cite{rafelski}) will 
not be washed out in the rescattering process and might be observable.

The  charged particle abundancies, $\approx 880$ at midrapidity, are in the 
middle of the expected multiplicity at $y=0$ which reaches from 600 
to 1200 \cite{wang,qm99pred}. 
It is interesting to note that the total particle yield is
similar to the RQMD results \cite{sorgeqm99} and also similar to a corrected 
pQCD based parton cascade model \cite{bass9908014}. 
However, note the qualitative difference
between the present results and those of a 
pQCD based transport approach \cite{bass_vni}:
the UrQMD calculations indicate a 
net-proton density of approx. 12 around midrapidity, whereas
the pQCD based approach predicts a net-proton density of only 3.
This large difference should allow experiments to discriminate 
those models.

Let us now turn to the transverse expansion dynamics:
since the early UrQMD dynamics is based on string degrees of freedom,
newly created quarks are not allowed to interact until they have
finished their coalescence into hadrons (typically this requires 1fm/c in
the local restframe of the coalescing quarks).  
Due to the large Lorentz $\gamma$-factor, this leads to a relatively
small pressure in the initial reaction phase as compared to an 
equation of state which includes a phase transition to a thermalized QGP. 

\begin{figure}[t]
\vskip 0mm
%\vspace{-1.8cm}
\centerline{\psfig{figure=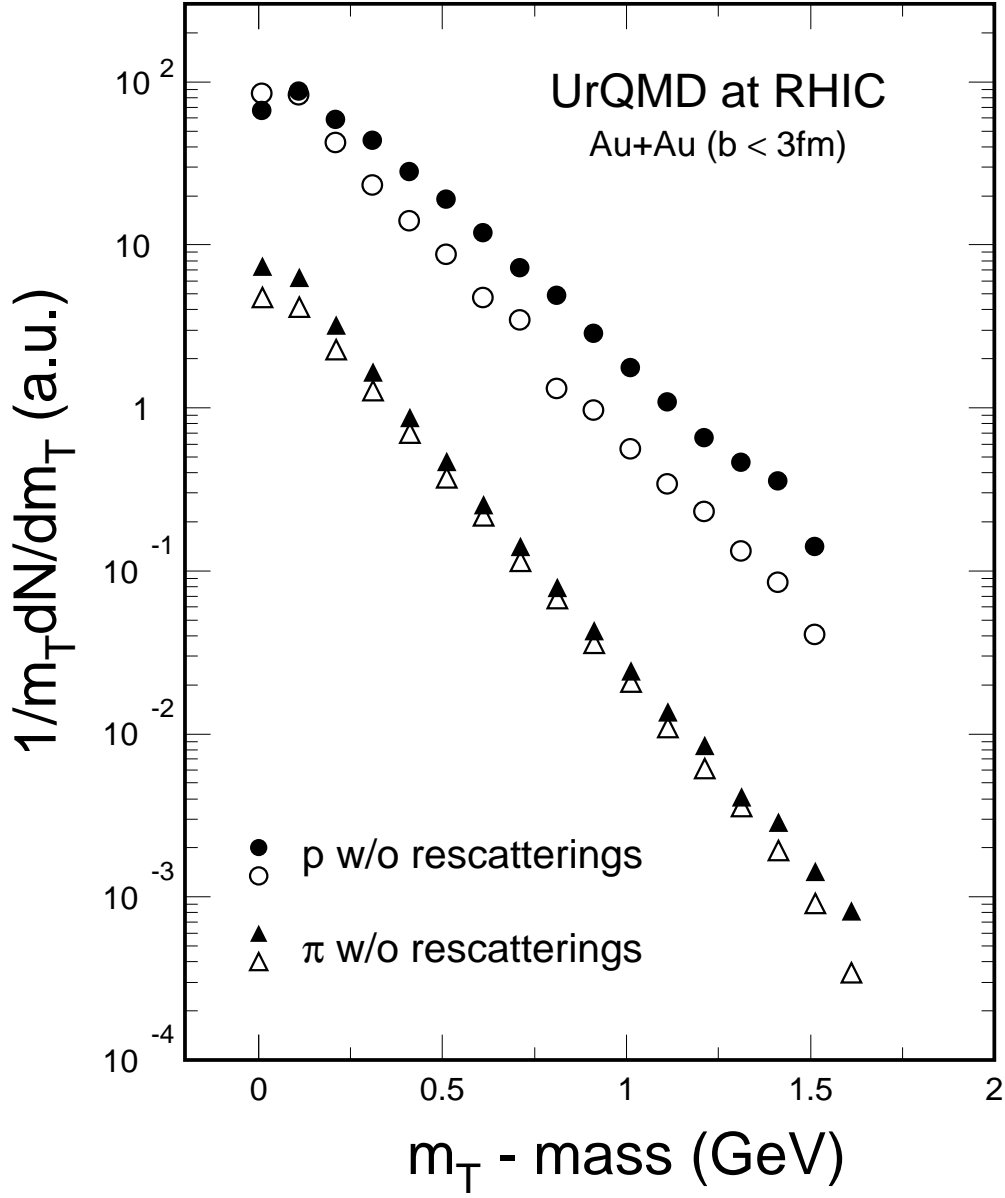,width=17cm}}
\vskip 0mm
\vspace{-.0cm}
\caption{Transverse mass distribution of protons (circles) and pions
(triangles) 
at midrapidity ($|y|<0.5$) in  Au+Au, $\sqrt{s}=200$~AGeV, b$<$3~fm. Full
symbols denote calculations with full rescattering. Open symbols denote 
calculations without meson-meson and meson-baryon interactions.\label{mt}}
\end{figure}

This behaviour is clearly visible in Fig. \ref{mt}: 
the transverse mass distribution 
of protons (circles) and pions (triangles) at midrapidity ($|y|<0.5$) 
are depicted for Au+Au, $\sqrt{s}=200$~AGeV, b$<$3~fm reactions. Full
symbols denote calculations with full rescattering and open symbols denote 
calculations without secondary interactions. Without rescattering the inverse
slopes of pions (open triangles)  and protons (open circles) are similar.
With full rescattering (full symbols) one observes a splitting 
in the inverse slopes of pions and protons. Hence, secondary
scatterings clearly create additional transverse flow.
Note that this effect is not visible in the pion spectrum: The pion 
slope differences with and without rescattering are marginal. 
The total number of pions is decreased without
rescattering. 

\begin{figure}[t]
\vskip 0mm
\vspace{-1.8cm}
\centerline{\psfig{figure=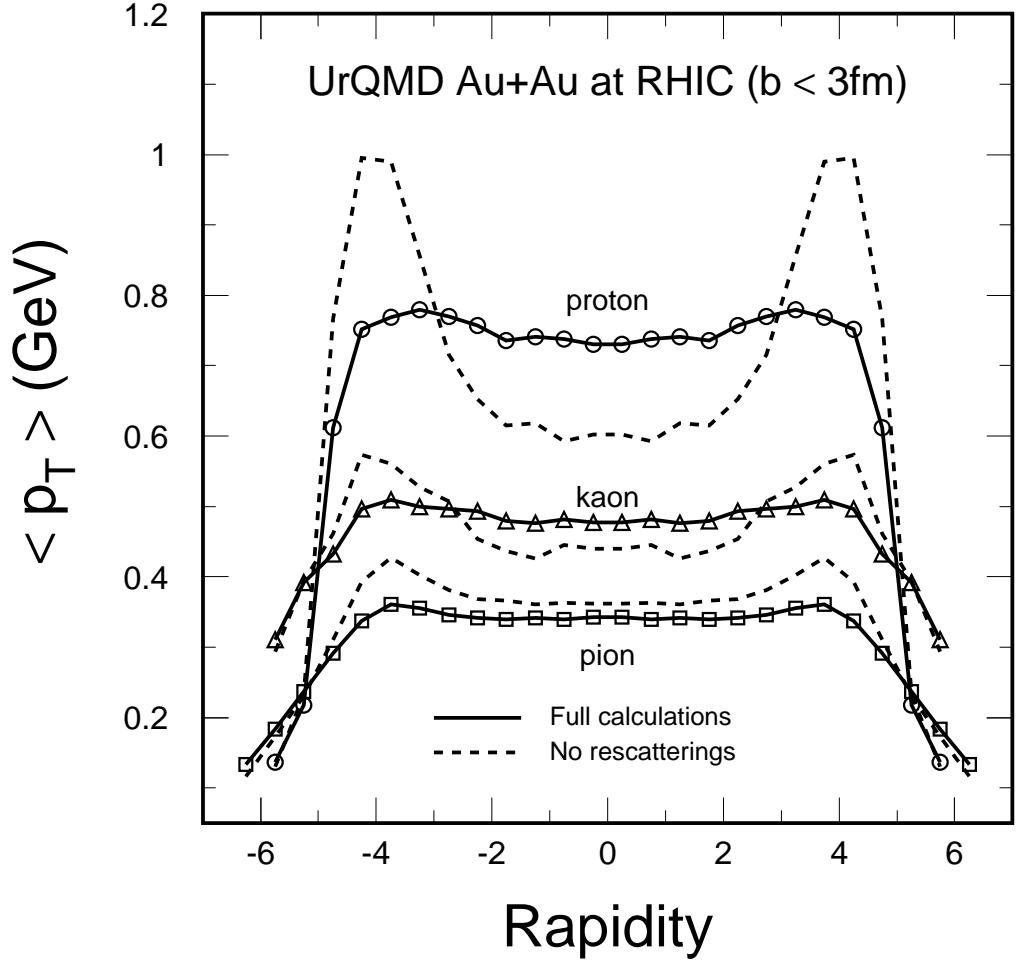,width=17cm}}
\vskip 0mm
\vspace{-.0cm}
\caption{Mean transverse momenta of protons, kaons and pions as a 
function of rapidity in  Au+Au, $\sqrt{s}=200$~AGeV, b$<$3~fm. 
Full lines denote
calculations with full rescattering. Dotted lines denote
calculations without meson-meson and meson-baryon interactions.\label{mpt}}
\end{figure}

This observation is supported by the mean transverse momenta of protons, kaons 
and pions, which are shown in Fig. \ref{mpt}  as a function of 
rapidity. 
Full lines denote calculations with full rescattering, whereas dotted 
lines denote calculations without rescattering. 
Without secondary interactions protons and pions
show mean transverse momenta at central rapidities similar 
to the values observed in $pp$ collisions. However, the mean 
transverse momenta rise strongly towards the target and projectile
region. This effect is known from pp collisions as the 'sea-gull' 
effect \footnote{This 'sea-gull'-feature
can be seen, if one plots the Feynman $x_F$ distribution instead of the
rapidity distribution.} \cite{seagull}. 
Frequent rescattering leads
to a hydrodynamic type behaviour - this is
demonstrated in Figs. \ref{mt} and \ref{mpt}. 
The mean $p_T$ differences between proton and kaon become much
larger than the difference between pion and kaon, a characteristic sign of
hydrodynamic flow\footnote{Protons always show some flow, due to
the Baryon-Baryon interactions. If the model is run in pure first collision
mode, this difference vanishes.}. 

With full rescattering the mean transverse momenta of protons increase at 
central rapidities and decrease in
the target/projectile region. This  leads to a flat mean $p_{\perp}$ over 
rapidity. The same effect works for the kaons. 
In contrast, pions cool down
due to rescattering (compare the dotted and full lines for pions). 
The cool off of pions is due to: (i) s-channel $\pi +p$ interactions which
result in a splitting of pions and proton slopes due to the decay kinematics of
the baryon resonances; (ii) inelastic interactions of pions lead to a
destruction of the pions with high transverse momenta (high energy), 
e.g. $\pi + \pi \rightarrow K\overline K$; (ii) it is in general 
difficult to heat pions up to more than 140~MeV, because of 
production of new pions above this temperature.
Pions lose part of their kinetic  energy to create new 
hadrons and by pushing the surrounding baryons, kaons, etc. aside. 
Thus, the pions act as an energy reservoir for the 
heavier hadrons (pion wind \cite{oscar}).

In conclusion, the UrQMD model has been applied to
Au+Au reactions at RHIC energies. This model treats the dynamics of the hot and
dense system by constituent (di-)quark and hadronic degrees of freedom. The
collision spectra have been studied and the effects of secondary interactions
have been quantified.  Substantial baryon stopping power has been predicted. 
The resulting particle production has been analyzed.  Secondary interactions
are found to be very important for such a strong baryon stopping. They
constitute a sizable source for particle production.
The study of the transverse expansion of the system revealed that it 
is driven by pions (''pion wind''): Pions 
transfer their energy in the expansion phase to the heavier hadrons. As a 
result the pions are cooled as rescattering is included.  
The overall particle production is found to be similar to 
pQCD motivated models. However, the net-proton rapidity density at
$y_{c.m.}$ differs by a factor of 4 between both approaches.
This may be used to experimentally distinguish between these models.

\section*{Acknowledgements}

This research used resources of the
National Energy Research Scientific Computing Center (NERSC).
This work is supported by the U.S. Department of Energy under contract
No. DE-AC03-76SF00098, the BMBF, GSI, DFG and Graduiertenkolleg
'Theoretische und experimentelle Schwerionenphysik'. 
S.A. Bass is also partially supported by the DOE grant DE-FG02-96ER40945.
M. Bleicher wants to thank the Nuclear Theory Group at LBNL for support
and fruitful discussions.

\newpage
\begin{table}[t]
\begin{tabular}{l|c|c}
Reaction           &  with rescattering  & without rescattering\\\hline
Baryon--Baryon     & yes  & yes  \\
Baryon--Diquark   & yes  & yes  \\
Baryon--Quark      & yes  & no   \\
Meson--Baryon      & yes  & no   \\
Meson--Diquark    & yes  & no   \\
Meson--Quark       & yes  & no   \\
Meson--Meson       & yes  & no   \\
Diquark--Diquark & yes  & yes  \\
Quark--Diquark    & yes  & no   \\
Quark-Quark        & yes  & no   \\
Anti-Baryon--Baryon       & yes  & yes  \\
Anti-Diquark--Diquark   & yes  & yes  \\
Quark--Anti-Diquark      & yes  & no   \\
Anti-Quark-Anti-Quark     & yes  & no   \\\hline
\end{tabular}
\caption{Possible Reaction channels with and without rescattering. Quark and
Diquark refer to constituent Quarks and Diquarks at the string end
points. If not especially mentioned, antiparticle reactions behave like
particle reactions.\label{tb1}}
\end{table}

\newpage

\newpage

\newpage

\newpage

\newpage

\end{document}